# The orbital phases and secondary transit of Kepler-10b

# -  A physical interpretation based on the *Lava-ocean planet* model  -

Authors: D. Rouan[a], H.J. Deeg[b,c], O. Demangeon[d,e], B. Samuel [d,e], C. Cavarroc [d,e] , B. Fegley[f], A. Léger[d,e].

*Affiliations*

a. LESIA, UMR 8109 CNRS, Observatoire de Paris, UVSQ, Université Paris-Diderot, 5 pl. J. Janssen, 92195 Meudon, France; e-mail:  daniel.rouan@obspm.fr

b. Instituto de Astrofísica de Canarias, E-38205 La Laguna, Tenerife, Spain

c. Universidad de La Laguna, Dept. de Astrofísica, E-38200 La Laguna, Tenerife, Spain

d.  Institut  d'Astrophysique  Spatiale,  Université  Paris-Sud,  bât  121,  Univ.  Paris-Sud,  F-91405 Orsay, France

e. Institut d'Astrophysique Spatiale, CNRS (UMR 8617), bât 121, Univ. Paris-Sud, F-91405 Orsay, France

f. Planetary Chemistry Laboratory McDonnell Center for the Space Sciences, Dep. of Earth and Dept. Planetary Sciences, Washington University in St. Louis, USA

## Abstract

The *Kepler* mission has made an important observation, the first detection of photons from a terrestrial planet by observing its phase curve (Kepler-10b). This opens a new field in exoplanet science: the possibility to get information about the atmosphere and surface of rocky planets, objects of prime interest.

In this letter, we apply the *Lava-ocean* model to interpret the observed phase curve. The model, a planet with no atmosphere and a surface partially made of molten rocks, has been proposed for planets of the class of CoRoT-7b, i.e. rocky planets very close to their star (at few stellar radii). Kepler-10b is a typical member of this family. It predicts that the light from the planet has an important emission component in addition to the reflected one, even in the *Kepler* spectral band. Assuming an isotropical reflection of light by the planetary surface (Lambertian-like approximation), we find that a Bond albedo of ~50% can account for the observed amplitude of the phase curve, as opposed to a first attempt where an unusually high value was found. We propose a physical process to explain this still large value of the albedo.

The overall interpretation can be tested in the future with instruments as JWST or EChO. Our model predicts a spectral dependence that is clearly distinguishable from that of purely reflected light, and from that of a planet at a uniform temperature.

Short title: *Lava-ocean* model applied to Kepler-10b

Text

## Introduction

The *Kepler* mission (Koch et al. 2010) has recently discovered a new super-Earth, Kepler-10b (Batalha et al. 2011, thereafter B+2011). It is a member of what looks as a new family, the CoRoT-7b-like planets or *Lava-ocean* planets (Léger, Rouan, Schneider et al. 2009; Léger et al. 2011, thereafter L+2011).  Presently, this family contains two members, CoRoT-7b and Kepler-10b.

These planets share several key properties: (i) a radius less than 2 Earth radii, a mass less than 10 Earth masses; (ii) measured mass and radius pointing to a rocky composition; (iii) a very short orbital period, and consequently a proximity to their parent star ($P$=0.85, and 0.84 day; $a$=0.0171, and 0.0168 AU; for CoRoT-7b and Kepler-10b, respectively). Their central stars are solar-like.

The planet proximity to the central star likely induces a phase locking of spin and orbital rotations (locking time <1 Myr).  Consequently, the dayside continuously faces the star, and the nightside is in the dark, which produces a major asymmetry between them. L+2011 proposed a physical model for such planets predicting several specific features, including a temperature map resulting only from a local equilibrium between the emitted radiation, and the received radiation (dayside) or geothermal flux (nightside).  They derived high surface temperatures, 2500 and 3040 K at the respective sub-stellar points, which results in the fusion and vapor fractionation of rocks. This produces a lava-ocean made of refractory components extending over a significant fraction of the dayside (zenith angles ≤51° and ≤74°, respectively).  The night side is very cold (50-75 K), setting the dichotomy between the bright and dark faces into numbers.

The extreme accuracy of the *Kepler* photometry has permitted *the first direct detection of photons from a terrestrial planet,* by observing changes in the system flux as a function of the planetary phase, and possibly detecting the secondary (back) transit (B+2011).  It is remarkable that *Kepler* has succeeded in detecting this phase curve of a terrestrial exoplanet right at the beginning of the mission. The only phase curves detected so far were those of giant exoplanets (Cowan & Agol 2011) that produce a much larger signal due to their size. This detection opens a new field because it is the only tool we will have in the near future to investigate atmospheres and surfaces of terrestrial planets.

B+2011 fitted the folded light curve of Kepler-10b with two parameters: the phase curve amplitude and the secondary transit amplitude. They found $(\Delta F/F)_{phase}$=7.6±2.0 ppm (1ppm≡$10^{-6}$), and $(\Delta F/F)_{sec-trans}$=5.8±2.5 ppm. These values, with their uncertainties, are compatible with the expectation that the phase curve amplitude is smaller than, or equal to, the secondary transit amplitude when complete occultation occurs. In a first attempt, B+2011 proposed an interpretation in terms of purely reflected light but found an unusually high geometric albedo, $p$=0.68, corresponding to a Bond albedo $A_B$=1.02±0.27 in the case of isotropical (Lambertian-like) reflection properties ($p$=(2/3)$A_B$, Seager et al. 2000). They concluded that the emission of the planet had to be taken into account but did not estimate it.

Here, we reanalyze the same data but with a single parameter, and propose a physical and self-consistent interpretation, within the *Lava-ocean* model (L+2011). We propose a test of this interpretation, measurements in the near-IR where the emission dominates, using space instruments as JWST and/or EChO [1].

## Light components from Keper-10b

Immediately before and after a primary transit the radiation from a planet comes essentially from its nightside, and before and after the secondary transit from its dayside. They have two components: (i) the planetary thermal emission, a function of the temperature and emissivity across the planet; (ii) the reflected starlight. In the following, we estimate the total amplitude of the planetary phase curve from extrapolations of the measured curve as if there were no transits.

## Thermal emission

---

[1] The Exoplanet Characterization Observatory (EChO) is a proposed mission to investigate exoplanetary atmospheres, presently studied by ESA (Tinetti et al. 2011).

The nightside is so cold (50 – 75 K) that its contribution in the visible is negligible. On the other hand, the dayside is heated up to temperatures over 3000 K (L+2011) and emits in the spectral domain where *Kepler* is sensitive (Koch et al. 2010). The *Lava-ocean* planet model produces a temperature distribution where $T$ is a function of the stellar zenith-angle, $\theta$, only (cylindrical symmetry). For the hot regions, it is (Léger, Rouan, Schneider et al. 2009):

$$T_{pl}(\theta)=(\varepsilon_5/\varepsilon_3)^{1/4}.(R_{st}/a)^{1/2}.[\cos(\theta)]^{1/4}.T_{st}, \qquad (Eq.1)$$

where $\varepsilon_{3,5}$ are the mean emissivities (absorptivities) of the planetary surface in the wavelength domains corresponding to the planetary emission ($T_3\sim 3000$ K), and stellar emission ($T_5\sim 5000$ K); $R_{st}$ and $T_{st}$ the stellar radius and effective temperature; $a$ the orbital semi-major axis. Thermodynamics implies $\varepsilon(\lambda)=1-A_B(\lambda)$ (Kirchhoff's law).

Petrov and Vorobyev (2005) have performed precise laboratory measurements of the emissivity of liquid alumina ($Al_2O_3$), a proxy for the composition of the ocean proposed by L+2011, $(Al_2O_3)_{0.87}(CaO)_{0.13}$. They found $\varepsilon(0.6\mu m)/\varepsilon(1.3\mu m)=1.04$. We will thus assume that the emissivity of the planetary surface is independent of wavelength (grey approximation), and denote it $\varepsilon$. Using $a/R_{st}=3.436$ and $T_{st}=5627$ K (B+2011), the temperature map of Kepler-10b becomes:

$$T_{pl}(\theta)=3040.[\cos(\theta)]^{1/4} \quad (K). \qquad (Eq.2)$$

The photon spectral density of an emitting surface is $\varepsilon.\mathbb{N}_P[\lambda, T(\theta)]$, where $\mathbb{N}_P$ is the Planck function in photon number. The planetary emission near the secondary transit, and that of a GV star in the effective temperature approximation, both into the direction of the observer, are:

$$N_{pl}(\lambda) = 2\pi R_{pl}^2.\varepsilon. \int_0^{\pi/2} \mathbb{N}_P[\lambda, T(\theta)]\cos(\theta)\sin(\theta)\,d\theta \qquad (Eq.3)$$

$$N_{st}(\lambda) = \pi R_{st}^2.\mathbb{N}_P[\lambda, T_{st}]. \qquad (Eq.4)$$

Figure_1

projected molten surface dominates (93% of the total).

Specular reflection contribution can be estimated assuming that the planetary surface is a liquid dielectric dioptre. In the Gaussian approximation, at the ''full planet'' phase, the stellar image in the planetary spherical mirror has a radius $R_{st}R_{pl}/2a$, and the image/star ratio of received lights is $(r_{sp}/4)(R_{pl}/a)^2$, with $r_{sp}$ the Fresnel reflection coefficient. For liquid alumina ($n(0.6\mu m)=1.74$, Krishnan et al. 1991) and normal reflection, $r_{sp}$ is 7.7%. Then, specular reflection alone provides quite a small contribution, 0.019 ppm. Although it can be observed in cases where an image is possible, e.g. for Earth (Williams & Gaidos 2008), it is two orders of magnitude lower than other contributions (Eq.6), and is negligible.

Near the secondary transit, ratio of reflected to stellar lights is $r_{refl} = p(R_{pl}/a)^2$. For an isotropical spherical reflector (Lambert-like approximation), the reflected part of the phase curve is (Seager et al. 2000):

$$r_{\text{refl}}=(2/3)A_B(R_{pl}/a)^2,$$

or, using $R_{pl}$ and $a$ values from B+2011:

$$r_{\text{refl}}=8.58A_B \quad \text{(ppm)}. \tag{Eq.7}$$

The secondary transit is the sum of its emitted and reflected planetary components, $n_{\text{tot}}=r_{\text{emis}}+r_{\text{refl}}$. $\varepsilon=1-A_B$ and Eq.6-7 give:

$$n_{\text{tot}}=2.85(1-A_B)+8.58A_B \quad \text{(ppm)}. \tag{Eq.8}$$

**Fitted amplitude of the orbital phase curve - resulting albedo -**

Observations by B+2011 are shown in Fig.2. The S/N is rather low, as expected for such a difficult observation. We do not try a fit without presuppositions but assume an *a priori* knowledge of the process and fit its amplitude.

In the *Lava-ocean planet* model, atmosphere is made of rock vapors and is tenuous ($P$<1mbar) so that winds cannot play a significant role in the temperature distribution. Planetary emission and reflection towards the observer are maximal at the "full planet" and vanishing at the "new planet" (Fig.1). The secondary transit starts close to the phase curve maximum and its depth equals the phase curve amplitude.

The situation could be different for a model implying a thick atmosphere able to reduce the contrast between day and night. This would lead to some thermal emission from the nightside around the ''new planet'', and reduce the phase curve amplitude. The secondary transit (no emission) would be "deeper" than its minimum.

In our interpretation, the observed phase curve and secondary transit can be fitted with a *single parameter*, the amplitude of the sine curve approaching the former (Samuel 2011). A least mean square fit yields a ptv amplitude (Fig.2):

$r_{fit}$=5.6±2.0   (ppm).                                                  (Eq.9)

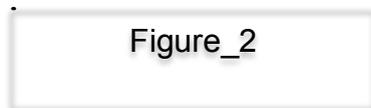

Relation $r_{fit}$=$r_{tot}$ and Eq.8-9 imply

$A_B$=0.48±0.35.                                                          (Eq.10)

This value is significantly less than that proposed by B+2011 (1.02±0.27) for two reasons: (i) our fit points to a smaller phase curve amplitude (5.6 vs 7.6 ppm), (ii) the thermal emission contributes about 1/3 to the total (Fig.3_left). B+2011 were aware of the problem and considered their unusual high value as an indication that reflection alone failed explaining satisfactorily the observations.

'              Figure_3

## Possible origin of the high albedo value

A ~0.50 Bond albedo is still pretty high for an atmosphere-less planet, even if estimated with a large uncertainty.  In particular, it is much larger than expected from pure Fresnel reflection (7.7% for $Al_2O_3$ at normal incidence).  This lava is almost as white as snow, and very different from that of terrestrial volcanoes.  Is this realistic?

On Earth, white materials are usually made of inclusions of a non-absorbing material, with size similar to the light wavelength, within a non-absorbing matrix, e.g. water clouds, snow, white waters, white quartz, milks, white paints, paper...  In this spirit, we propose two possible and complementary phenomena.

First, a low absorption matrix is expected.  Electronic transitions of unpaired d-electrons of transition metal ions in different crystallographic sites are responsible for the colors of terrestrial minerals and lavas (Burns 1993).  The ions are in crystal fields removing the degeneracy of the d-orbitals and splitting them into levels with separations corresponding to visible light.  In the proposed $Al_2O_3$-CaO melt, calcium and aluminum ions have no unfilled d-orbitals and cannot absorb visible light.

Second, inhomogeneities within the lava can backscatter efficiently incident light. This situation is found in Earth's oceans ($A_B$=6%, $r_{sp}$=2%) with backscattering from colloidal suspensions, microbic algae, plankton… that also gives the different colors of seawater. In the case of *Lava-ocean* planets, what could be the floating solid particles of non-absorbing compound(s)?

The candidate must melt above 3040 K, be insoluble in the $Al_2O_3$-CaO liquid, form small particles to efficiently scatter visible light, and be abundant enough. The most refractory oxides are: MgO ($T_{fus}$=3105 K), CaO (3200 K), $UO_2$ (3140 K), and $ThO_2$ (3660 K) (JANAF 1985). The first two are soluble in the melt, but the last two could be insoluble and precipitate (Mao et al. 2004). The abundances of $UO_2$ and $ThO_2$ (thoria), relative to alumina (Lodders et al. 2009), are $2.1 \ 10^{-7}$ and $1.1 \ 10^{-6}$ by volume. Thoria is the most refractory and abundant candidate. Vapor fractionation should increase its abundance by ~10 due to the ratio of $Al_2O_3$ to $ThO_2$ vapor pressures (Ackermann et al. 1963; Fegley and Cameron 1987), leading to a mixing ratio of ~$10^{-5}$.

Mie calculation of light scattering by dielectric spheres (Hulst 1981) gives a rather isotropic extinction coefficient, $Q_{ext}$, proportional to radius $r$ (absolute cross section $\propto r^3$), for $x$=2.3$r/\lambda$~1 ($Q_{ext}$~1), suitable to the Mie curve for $ThO_2$ refractive index ($n$=2.2) embedded in $Al_2O_3$ ($n$=1.7). If a significant fraction of $ThO_2$ particles have radii in this regime ($r$~0.1-0.4 $\mu$m), which is compatible with laboratory condensate sizes (Haggerty 1984), an unit optical depth would be achieved for ~10 cm of lava leading to a very efficient backscattering without major absorption, and then a high albedo.

We have no further indications that precipitate radii are in that range, but the above estimates indicate that *ThO_2 particles, dispersed[2] in the Al_2O_3-CaO lava,* can produce *a 50% Bond albedo.*

**Possible tests of the model**

Discriminating tests of our interpretation of observations can be foreseen in the mid-future when suitable space instruments will be available. According to the mean value of the inferred Bond albedo, ~50%, the bolometric power emitted by the planet is similar to the reflected one. Within the *Kepler* spectral band, reflected light is larger than emitted light (Fig.3_left) because the planet is a cooler source (≤3040 K) than the star (5600 K). This suggests a test using an instrument with a spectral response further in the IR, as JWST, or EChO. Fig.3_middle shows the photon density of emitted and reflected light in near-IR, and the corresponding signals in different JWST spectral bands. The component that dominates the planetary light is predicted to change near 1 μm, when going from the visible to longer wavelengths.

---

[2] The large density of $ThO_2$ (10 g/cm$^3$) would not lead to full particle sedimentation, in a way similar to water droplets in terrestrial white clouds because sedimentation time for clearing 1 m of lava [4$(r/0.1\mu m)^{-2}$ yrs, using Stokes drag ($F$=6$\pi\eta rv$) with viscosity $\eta$=0.04 Pa.s (L+2011) and the negative buoyancy of thoria in alumina] is comparable to the ocean turn over time (~2 yrs) due to the currents driven by the horizontal temperature gradient ($v$~0.2 m/s) with Coriolis forces (L+2011).

Fig.3_right shows the expected amplitude of the secondary transit signal in these bands. The *Lava-ocean* model predicts a spectral variation *significantly different from that of purely reflected light* (constant amplitude, for a grey albedo), *or from a planet whose temperature would be rendered uniform by a very thick atmosphere* ($T_{unif}$=2150 K, for $A_B$=0.5), which could be tested. The JWST integration times needed to obtain the proposed accuracy (6$\sigma$ at 4.4 μm) are long (200 h per filter), but other strategies can be considered.

Analogous observations could be performed on 55Cnc-e (Winn et al 2011) in much shorter integration times, typically by *two orders of magnitude* given the intensity ratio of Kepler-10 and 55Cnc stars (V=6 and V=11, respectively). The use of the JWST spectrometers would avoid detector saturation. This planet is not studied in this letter because, according to the information available today, the *Lava-ocean* planet model does not apply to it. However, such measurements would provide valuable clues on the physical processes occurring at the surface of that planet, whatever they are. They should be rather easy to obtain thanks to the high brightness of the stellar system, a real jewel that nature offers.

As *Kepler* keeps on observing, the S/N ratio of its measurements will improve (by a factor $5^{1/2} \sim 2.2$ at the end of the five year mission, assuming no major improvement in the data reduction), and should better constrain the physical interpretations. In particular, it should be possible to reduce the uncertainty on the inferred albedo.

Now, will it be possible to measure the phase curve amplitude and the secondary transit separately so that it can be decided whether they are equal, or not, and whether there is energy redistribution at the planetary surface? In the *Lava-ocean* interpretation, secondary transit depth and phase curve amplitudes are equal. The emission part of the phase curve amplitude is 1.4 ppm, reflection part is 4.2 ppm (Fig.3-left).

If the redistribution of energy was total, and the surface temperature uniform, the emission part of the phase curve would be null and the difference between the secondary transit and the phase curve amplitude would be the dayside emission. The bolometric value of the latter would be a quarter of the total for $A_B$=0.5, occurring in the near-IR (Fig.3-right). In the *Kepler* band, this dayside emission can be overestimated by its value in the *Lava-ocean* interpretation, 1.4 ppm, which should be compared to its expected uncertainty. The present value of the error on the secondary transit is 3.3 ppm (open circles in Fig.2), and should reduce to 1.5 ppm at the end of the mission. The error on its difference with the phase curve will be the quadratic sum of that on the secondary transit, and that on the phase curve (2.0/2.2=0.9 ppm), or 1.8 ppm ($1\sigma$). As this is larger than the quantity to be measured ($\leq$1.4 ppm), we conclude that *Kepler* will *not be able to determine whether there is energy redistribution at the surface of the planet*, unless a major improvement in the data reduction is achieved.

As stated, the situation would be different if instruments having spectroscopic capabilities and (very) high stability in the near-IR are built. For instance, observations with JWST should be able to discriminate between the different hypotheses, *Lava-ocean* model/pure reflection/total surface thermalisation, at the level of ~$6\sigma$ for integration times of the order of 200 h per filter (Fig.3-right), and may be even more efficient in pure spectroscopic modes.

In any case, the planetary phase observations by *Kepler* are steps towards the spectral study of terrestrial planets in the Habitable Zone of their star, objects of major interest.

**Acknowledgements**

The authors thank V.Petrov for valuable information on optical properties of molten alumina, F.Selsis, L.Léger, P.Bordé, P.Gaulme and M.Ollivier for important discussions, J.Leibacher for manuscript improvements, and an anonymous referee for a very valuable suggestion. AL, OD, BS, and CC thank CNRS-PNP, BF NSF Grant AST0707377, HD grants ESP2007-65480-C02-02 and AYA2010-20982-C02-02.

# FIGURES

Full planet

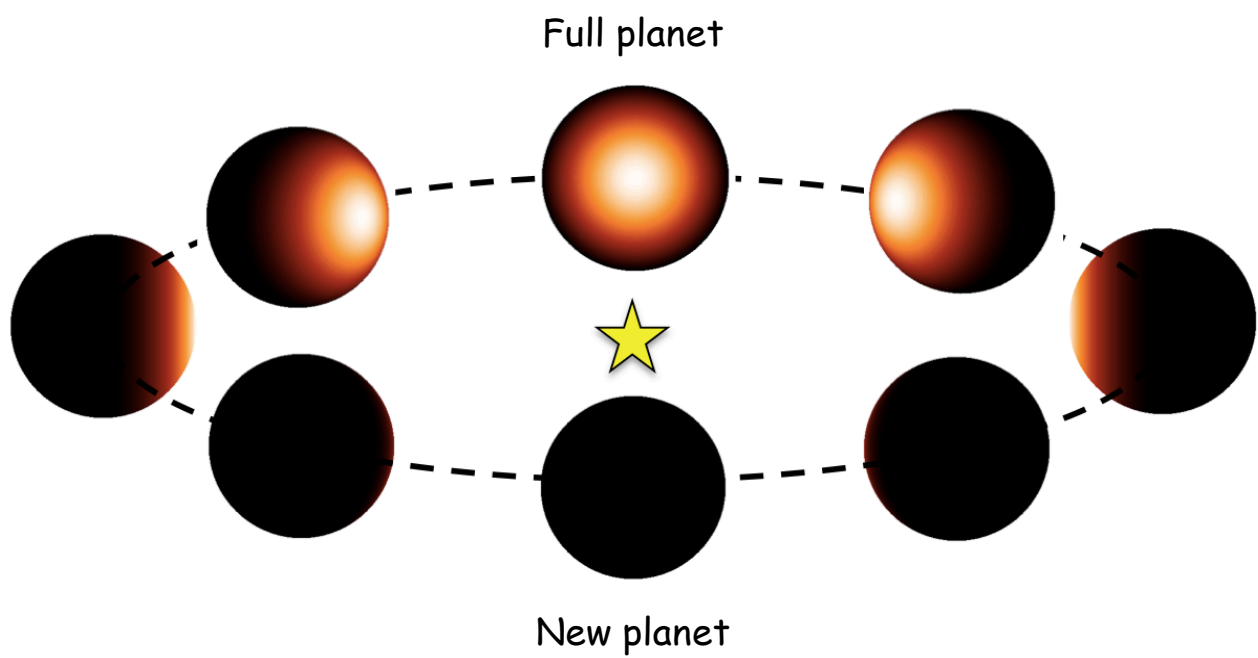

New planet

<u>Figure_1</u>:  Representation of emitted light from Keper-10b during its different phases, according to the *Lava-ocean* model.  The reflected light has a similar geometry, but a spectral content close to that of the stellar light.

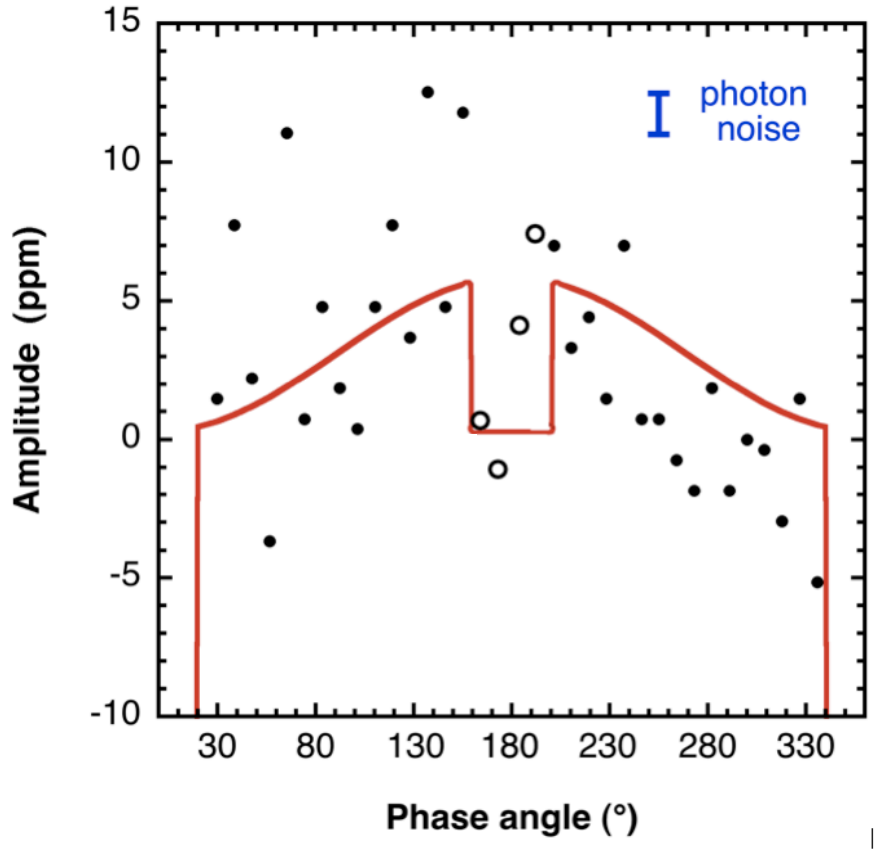

Figure_2: Folded light curve of Kepler-10b versus the planetary orbital phase (data from B+2011). The phase origin is deduced from the primary transit. The full line corresponds to a quadratic fit to a sine phase curve (Samuel 2011), assuming a circular orbit, and a secondary transit that has the same width, an almost instantaneous ingress/egress ($R_{pl} \ll R_{st}$), and a floor reaching the minimum of the orbital phase curve. These assumptions result in a single fitting parameter, the amplitude of the sine. Data points outside the secondary transit are represented as filled circles, and those inside as open circles, for the sake of clarity only. The value found for the total amplitude is $r_{fit} = 5.6 \pm 2.0$ ppm. The reported photon noise ($1\sigma$) is estimated from the flux of photoelectrons per phase bin. It is significantly less than the standard deviation of the data points, which indicates the presence of additional noise(s).

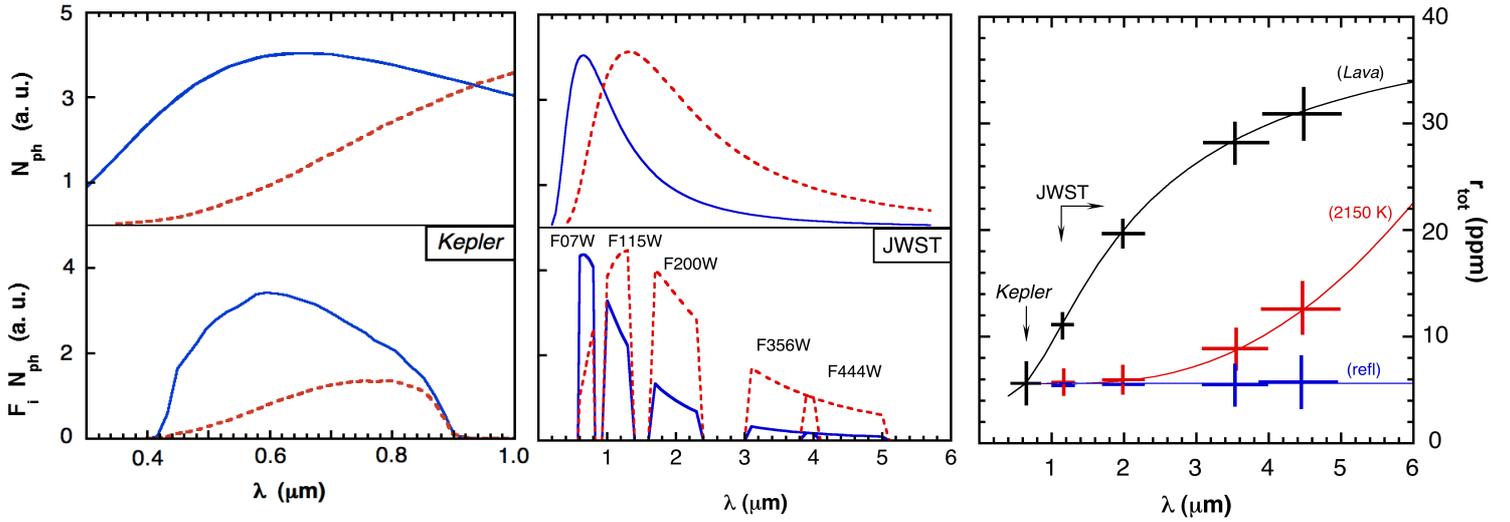

<u>Figure_3</u>.  <u>Left</u>, top panel: in the *Lava-ocean* model, spectral composition (photons) of the reflected component (solid line), and emission (dashed line) of light from Kepler-10b near the secondary transit, for a 0.5 Bond albedo.  Bottom panel: corresponding components of the *Kepler* signal (photo-electrons) after multiplying by the instrumental response, $F_i(\lambda)$.  <u>Middle</u>, top panel: same quantities (photons) but in a broader spectral range. Bottom panel: expectations in different JWST filters.  Components are estimated assuming a constant albedo from visible to near-IR (grey approximation).  As wavelength increases, the emission part of the signal increases relatively to the reflection part.  <u>Right</u>: expected amplitude of the secondary transit within different hypotheses.  Crosses labeled ''Lava'' correspond to the signals within the *Lava-ocean* model; crosses labeled ''refl'' to pure reflection of starlight; and those labeled ''2150 K'' to a planet at a uniform temperature of 2150 K (Eq.7).  Vertical bars are expected errors (1σ) for 200 hour long integrations per filter, as given by the JWST Exposure Time Calculator (2011) assuming 100 h spent on the secondary transit and 100 h out of it.  Observations by JWST should be able to *discriminate between these different interpretations.*